\documentclass[10pt]{article}
\usepackage{amsmath,amsfonts,amsthm,amssymb,amscd}

\newcommand{\be}{\begin{equation}}
\newcommand{\ee}{\end{equation}}
\newcommand{\bs}{\begin{split}}
\newcommand{\es}{\end{split}}
\newcommand{\ba}{\begin{align}}
\newcommand{\ea}{\end{align}}
\newcommand{\basl}[1]{\begin{align}\begin{split}\label{#1}}
\newcommand{\bas}{\begin{align}\begin{split}}
\newcommand{\bl}[1]{\begin{equation}\label{#1}}
\newcommand{\el}{\end{equation}}

\newcommand\R{\mathbb{R}}
\newcommand\C{\mathbb{C}}

\title{Quantum radiative corrections for a model  in NMR  \\ in  quantum electrodynamics  }

\author{ L. Amour, L. Jager, J. Nourrigat}

\date{Universit\'e de Reims, France}

\begin{document}

\maketitle

\begin{abstract}
In this article, we are interested in a spin model including the quantized electromagnetic field (photons). With this model of quantum electrodynamics (QED) related to nuclear magnetic resonance (NMR) we give  explicit quantum radiative corrections of the time evolution for the spin observables, for the electric and magnetic fields observables and for the photon number observable. As a by-product, this underlines that Bloch equations are the semiclassical limit of the model in QED considered here. In addition, transition probabilities for the same model are investigated.
\end{abstract}

\vskip 0.5cm

\parindent=0pt

{\it Keywords:} Quantum electrodynamics, Nuclear Magnetic Resonance, Interacting spins, Spin model, Radiative corrections, Photon emission, Photon number, Transition probabilities, Maxwell-Bloch equations, Bloch equations,  QED, NMR, Spin-boson model, Spin-photon model.

\

{\it MSC 2010:} 81V10, 81Q20.
\parindent = 0 cm

\parskip 10pt
\baselineskip 14pt

\section{Introduction}\label{s1}

Nuclear magnetic resonance consisting of an interaction between a system of $N$ spin $1/2$ fixed particles in  $\R^3$ and a constant magnetic field together with a plane electromagnetic wave is frequently modeled by the  Bloch equations  \cite{BL}. In this model, the particle spin of index $\lambda $
($1\leq \lambda \leq N$) is viewed as a vector ${\bf S}^{[\lambda ]} (t)$
in $\R^3$, solution to the following system, 
\be\label{9-4-1} \frac  {d } { dt} {\bf S}  ^{[\lambda  ]} (t) =
2 ( {\bf B} ^{ext}  + {\bf B} ^{[wave]}  (x_{\lambda} , t) )
\times  {\bf S}  ^{[\lambda  ]} (t), \ee
where $x_{\lambda}$ is the  point of  $\R^3$ where the particle $\lambda $
is located. Here ${\bf B} ^{ext}$ 
is the constant magnetic field and  ${\bf B} ^{[wave]}  (x , t) $ is the field associated with a plane electromagnetic wave.

If the plane wave is entirely vanishing and if the spin is initially pointing along the direction of the constant field, in one direction or the other, then it remains eternally fixed in this position according to Bloch equations. This seems somehow unrealistic and justifies to use a model in the QED setting which will show a  physically more consistent behavior in this case. Indeed, we shall explicitly show in Section \ref{s2} that, for the model coming from quantum electrodynamics (QED), the spin cannot remain eternally fixed in this configuration.

Another advantage of the QED model is a better description of the interactions between the particles being generally atomic nuclei. In particular, one could additionally include the role of the electrons  (see for instance \cite{P-T} and \cite{Ro-Au}  for this issue which also take into account relativistic effects) but it is not our case here.

We shall recall below the model coming from QED considered here being also described in Section 4.11
  of Reuse \cite{Reu}
  and we shall underline the role of the semiclassical parameter $\hslash>0$.
Then, we shall show that, when $\hslash$ goes to zero, the Bloch model is the semiclassical  limit of the QED model.
We shall present in a more explicit way the results of \cite{A-J-N} written in more abstract terms.
In Section \ref{smodel}, we shall introduce the model coming from QED.
 In Section \ref{s2}, we shall present a first quantum radiative correction  of order one  for the Bloch equations.   In Section \ref{s3}, we shall show that the photon average number variation between times $0$ and $t$ has a limit when $\hslash$ tends to zero. 
 In Section \ref{s4},  we shall estimate some transition probabilities.

\section{The model}\label{smodel}
 
{\it Spaces.} 

The Hilbert configuration space of photons 
is the space   $H$   of applications  $f \in L^2 (\R^3, \R^3)$ such that $k\cdot f(k) = 0$
almost everywhere in $k\in\R^3$ and its norm is defined by
$|f|^2=\int_{\R^3}|f(k)|^2dk$. The phase space of photons is $H^2$. It is frequently here identified
with the complexified $H_{\bf C}$.
The (real) scalar product of  $X$ and  $Y$ in
$H^2$ is denoted by $X \cdot Y$.
The Hilbert space ${\cal H}_{ph}$ of photon quantum states will be the
Fock space ${\cal F}_s (H_{\bf C})$.
The Hilbert space describing the states of $N$ fixed particles with spin $1/2$
without interaction at a given time is  ${\cal H}_{sp}  =( \C^2  )^{\otimes N}$.
The Hilbert space describing the states of the whole system is the completed tensor product
${\cal H}_{ph} \otimes {\cal H}_{sp} $.
We denote by  $< \cdot , \cdot >$
the scalar product in ${\cal H}_{ph}$ or in  ${\cal H}_{sp}$ or in the tensor product of these two spaces. It is ${\bf C}-$linear with respect to the right hand side.

{\it Operators.} 

Let  $M_{\omega}$ be the operator with domain   $D(M_{\omega}) \subset H$ such
that $M_{\omega} q (k) = |k| q(k)$ almost everywhere in $k\in\R^3$.
We denote likewise the analogous operators defined  on $H^2$ or
on the complexified $H_{\bf C}$.  Again, the spaces $H^2$ and
$H_{\bf C}$ are often identified throughout this paper. We shall use the photon Hamiltonian
$H_{ph} = \hslash d\Gamma (M_{\omega})$ and the number of photons operator
$N = d\Gamma (I)$ (see \cite{RSII} for the definition of $d\Gamma (\cdot)$).
At each point $x$ in $\R^3$ one defines below an element $ B_{jx}$ in  $H^2$ (when identifying $H^2$ with its complexified space $H_{\bf C}$),
\be\label{7.3} B_{jx}(k) = {i\chi(|k|)|k|^{1\over 2} \over (2\pi)^{3\over 2}}
e^{-i(   k\cdot x   )} {k\times  e_j \over |k|},\quad k\in\R^3\backslash\{0\}\ee
where $\chi $ is the ultraviolet cutoff and is a rapidly decaying function belonging to ${C} ^\infty(\R)$. With the above $B_{jx}$, we associate a mapping  $B_j (x , \cdot)$ defined on $H^2$ and an unbounded operator $B_j (x)$ in
${\cal H}_{ph}$ by
\be\label{def-champ} B_j (x , X) = B_{jx} \cdot X, \quad X\in H^2,\qquad
B_j (x) = \Phi_S ( B_{jx}), \ee
where $\cdot$ is the real scalar product on $H^2$ and $\Phi_S$
is the usual Segal field (see \cite{RSII}).
The operator   $B_j(x) = \Phi_S( B_{jx})$ corresponds to the  $j-$th component of the magnetic field at the point  $x$.
  We denote by
$J: H^2 \rightarrow H^2$  the helicity operator defined by, when $X=(q,p)$,
\be\label{7.5}J(X) (k) = \left ( {k\times  q(k) \over |k| } , {k\times  p(k) \over |k| } \right ),
\quad k \in \R^3 \setminus \{ 0 \}.\ee
At each point $x$ de $\R^3$,  we set $E_{jx} = JB_{jx} $ and we define a mapping $E_j (x , \cdot)$  on $H^2$ together with an unbounded operator
$E_j (x)$ in ${\cal H}_{ph}$ by,
\be\label{def-elec} E_j (x , X) = E_{jx}\cdot X = - B_{jx}\cdot JX,\qquad
E_j (x)= \Phi_S( E_{jx}). \ee
Thus,
$E_j(x) = \Phi_S( E_{jx})$ is the operator standing for the  $j-$th component of the electric field at $x$.
In the space  ${\cal H}_{sp} $, we shall use the operators associated with the
spins of the different particles. Let  $\sigma _j$ ($1 \leq j \leq 3$)
be the Pauli matrices,
\be\label{Pauli} \sigma_1 = \begin{pmatrix}  0 & 1 \\ 1 & 0   \end{pmatrix},
\qquad
\sigma_2 = \begin{pmatrix} 0 & -i \\ i & 0   \end{pmatrix},
\qquad
\sigma_3 = \begin{pmatrix}  1 & 0 \\ 0 & -1  \end{pmatrix}.\ee
For all $\lambda \leq N$ and all  $m\leq 3$, we denote by $\sigma_m^{[\lambda]}$
the operator in ${\cal H} _{sp}$ defined by,
\be\label{spin-ini}\sigma_m^{[\lambda]} = I \otimes \cdots \sigma_m \cdots \otimes I,
\ee
where  $\sigma_m$ is located at the  $\lambda ^{th}$ position.

{\it Hamiltonian.} 

 This Hamiltonian is often used for modelling NMR in quantum field theory (see  \cite{Ro-Au} and also Section 4.11 in \cite{Reu}).
The Hamiltonian  is a selfadjoint extension of the
following operator, initially defined  on a dense subspace of
${\cal H}_{ph} \otimes {\cal H}_{sp} $,
\be\label{7.1} H(\hslash) = H_{ph} \otimes I + \hslash H_{int},\ee
where $H_{ph} = \hslash d \Gamma (M_{\omega}) $ is the photon free energy operator  and
\be\label{7.8} H_{int} = \sum _{\lambda =1}^N  \sum _{m=1}^3 (B_m ^{ext} + B_m (x_{\lambda})) \otimes
\sigma_m^{[\lambda]},\ee
where  ${\bf B^{ext}}  = (B_1 ^{ext} , B_2 ^{ext} , B_3 ^{ext})$  is the constant external  magnetic
field and the  $x_{\lambda }$ ($1\leq \lambda \leq N$)  are the points of
$\R^3$ where the fixed particles are located.

{\it Evolution of observables.} 

The time evolution of a bounded or unbounded selfadjoint operator $A$  in ${\cal H}_{ph} \otimes {\cal H}_{sp} $ is defined by the operator,
\be\label{evol-A}
A (t, \hslash)=  
 e^{i {t \over \hslash}   H(\hslash) } 
 A
e^{-i{t \over \hslash} H(\hslash)}. \ee
The coherent states $\Psi_{X , \hslash}$  are defined for $X= (a , b)  $ in $H^2$
and  $\hslash>0$ by,
     $$ \Psi_{X \hslash} = e^{-{|X|^2  \over 4\hslash}} \sum _{n\geq 0} \frac
     {(a+ib) \otimes \cdots \otimes (a+ib)}  {(2\hslash)^{n/2} \sqrt {n!} } . $$
We call Wick (or alternatively Husimi) symbol of an operator  $A $ in ${\cal H}_{ph} \otimes {\cal H}_{sp}$
the  $A (\cdot , \hslash)$  defined on $H^2$ taking values in ${\cal L}({\cal H}_{sp})$ satisfying for all $X$ in $H^2$ and any $a$ and $b$ in ${\cal H}_{sp}$,
$$ <A(X , \hslash)a,b> = < A  ( \Psi_{X \hslash}  \otimes a) ,  ( \Psi_{X \hslash}  \otimes b) > .$$
If the  initial state is the coherent state  $ \Psi_{X \hslash} \otimes a $ with 
$a$ in ${\cal H}_{sp}$ having a unit norm, then the average of the observable $A$
at time $t$ is $ < A (t, X , \hslash) a , a > $
when denoting by $A(t , X , \hslash)$ the Wick symbol of $A(t, \hslash)$.
We also know that the Wick symbol of the Segal field   $\Phi_S(a) $
with $a$ in $H^2$ is the function $H^2 \ni X \rightarrow  a \cdot X$.

Let us underline at this stage that  each component of the evolution observable Wick symbol for the spin, for the electric and magnetic field, that we respectively denote by $S_j^{[\lambda]}(t,X,\hslash)$, $E_j (x ,X,t,\hslash)$ and $B_j (x ,X, t,\hslash)$, is an operator in $ {\cal H}_{sp}$ (for fixed $j,X, t,\hslash,\lambda,x$). Then, the evolution observable Wick symbol for the spin, the electric and magnetic field, denoted respectively by ${\bf S}  ^{[\lambda ]} (t, X,\hslash )$, ${\bf E}(x ,X,t,\hslash)$ and ${\bf B} (x ,X, t,\hslash)$, is a triplet of operators in $ {\cal H}_{sp}$ (for given $X, t,\hslash,\lambda,x$).

{\it   Free evolution.}

For any operator (observable) $A$ in ${\cal H}_{ph} \otimes {\cal H}_{sp}$ we set,
$$A^{free} (t , \hslash) =  e^{i{t \over \hslash}  ( H_{ph} \otimes I) } A
e^{-i{t \over \hslash}  ( H_{ph} \otimes I)}.$$
The Wick symbol $A^{free} (t , X , \hslash)$ of the operator $A^{free} (t , \hslash)$
satisfies,
\be\label{chi-t} A^{free} (t , X , \hslash)= A ( \chi_t X , \hslash),\qquad
( \chi_t X) (k) = e^{-it|k|} X(k). \ee
The mapping $A^{free}(t , X , \hslash)$  is denoted by $B_j^{free} (x, t , X)$ or
$E_j^{free} (x, t , X)$ when  $A$ is one the operators  $B_j (x)$ or $E_j (x)$  ($x\in \R^3$) respectively.
Recall that, for all  $X$ in $H^2$, the mappings  $B_j^{free} (x, t , X)$ and  $E_j^{free} (x, t , X)$ satisfy Maxwell equations in vacuum with initial data $B_j (x , X)$ and $E_j (x,  X)$.
One sees, again identifying  $H^2$ and $H_{\bf C}$ that,
\be\label{Bjxt} B_j^{free} (x, t , X) = B_{j, x, t} \cdot X,\qquad B_{j, x, t} (k)  = B_{jxt}(k)= e^{it|k|}B_{jx}(k) ,\ee
where $B_{jx}$ is defined in (\ref{7.3}).

Our purpose here is to study an asymptotic expansion in powers of $\hslash$ for the average of some specific observables $A$  at time $t$:  the spin components, the electric and magnetic field components at any point
$x$ of $\R^3$ and the photon number. We  shall systematically prove that the evolution of the first term of the asymptotic expansion is related to the Bloch equations which may therefore be regarded as a semiclassical limit of the model coming from QED. Finally, an investigation of the propagator  $e^{-i{t \over h} H(\hslash)}$
enables us to estimate some transition probabilities.
Most of these results rely on a mathematical basis developed in  \cite{A-J-N-1}, \cite{A-J-N}, \cite{A-L-N-1}, \cite{A-L-N-2}
 and  \cite{J}.

\section{Spins and electromagnetic fields}\label{s2}

The observable Wick symbols $ A(t, X , \hslash)$ under consideration here are differentiable and even analytical on the photon phase space  $H^2$. It takes values in 
 ${\cal L}( {\cal H}_{sp})$.
 We are able to give a bound for the successive orders differential maps.
 For instance, if the initial observable is $A = I \otimes \sigma_m^{[\lambda]}$ (with $1 \leq m \leq 3$) then the Wick symbol of the same observable  at time  $t$ is denoted $S_m  ^{[\lambda  ]} (t, X, \hslash)$ and with these vectorial notations we show that,
\be\label{1.1}|(d^m  {\bf S}  ^{[\lambda  ]} (t, X , \hslash) )  (V_1 , \dots , V_m ) | \leq
(4 Q_t( V_1)) ^{1/2} \dots  (4 Q_t( V_m) ) ^{1/2},
\ee
where we have set,
\be\label{def-2-Qt} Q_t  (V) = 2^N |t| \sum _{\lambda = 1}^N \sum _{j=1}^3 \int _0^t
| ( B_j ^{free} ( x_{\lambda}, s, V) |^2 ds. \ee
The bound (\ref{1.1})  comes from  Theorem 7.2 of \cite{A-L-N-1} and from
Theorem 1.4 (point iv) of \cite{A-J-N}.
In the aim to express  (\ref{1.1}) nearby estimates, we say that a mapping  
 $F$
defined on $H^2$ and taking values in ${\cal L} ( {\cal H} _{sp})$ belongs to the set $S(H^2 , K Q_t)$ if,
\be\label{1.bis}|(d^m  F( X, \hslash) )  (V_1 , \dots , V_m ) | \leq C(F)
(K Q_t( V_1)) ^{1/2} \dots  (K Q_t( V_m) ) ^{1/2}.
\ee
The smallest constant $C(F)$ satisfying inequalities (\ref{1.bis}) is called the $S(H^2 , K Q_t)$ norm of $F$ .

We prove in 
\cite{A-J-N} (Theorem 1.7) that there exists a sequence of functions ${\bf S}  ^{[\lambda , j ]} (t, q , p)$ being explicitly computable and there is for any each $M$ a function  $R_M (t, q, p, h)$ satisfying,
$$   {\bf S}  ^{[\lambda  ]} (t, X, \hslash) =
 \sum _{j= 0}^M  \hslash^j {\bf S}  ^{[\lambda , j ]} (t, X)+
\hslash^{M+1} R_{M, S } (t, X \hslash).$$
In addition,  each ${\bf S}  ^{[\lambda , j ]} (t, \cdot )$ belongs to $S(H^2, 16^{j+1} Q_t)$ and    $R_M (t, \cdot, \hslash)$ belongs to $S(H^2, 16^{M+5} Q_t)$, with a norm
bounded independently of $t$ and $\hslash$, when $t$ remains in a compact subset of $\R$ and $\hslash$
varies in  $(0, 1)$.  In \cite{A-J-N}, the bound on the remainder term $R_M (t, q, p, h)$ is obtained with anti-Wick symbolic calculus when using the class of functions  $S(H^2, K Q_t)$.
The article \cite{A-J-N} provides two ways for computing the
${\bf S}  ^{[\lambda , j ]} (t, \cdot )$.  The computation of the  ${\bf S}  ^{[\lambda , j ]} (t, \cdot )$ can be effectuated either using  Heisenberg type's equations or using operator-valued differential equations found in Spohn \cite{SP}.  These computations are  then closely related to the corresponding computations in the case of  the magnetic and electric fields observables.

We indeed prove that there exist vector fields ${\bf B}  ^{[ j ]} (x, t, X)$ depending on $X\in H^2$ and taking values in  ${\cal L}({\cal H}_{sp})$ with the following asymptotic expansion,
$$   {\bf B}  (x, t, X, \hslash) =
 \sum _{j= 0}^M  \hslash^j {\bf B}  ^{[ j ]} (x, t, X)+
\hslash^{M+1} {\bf R}_{M, B } (x, t, X, \hslash). $$
The same expansion holds true for the electric field. The mapping  ${\bf R}_{M, B }$ and its counterpart
${\bf R}_{M, E }$ share the same properties as for $R_{M, S}$.

Let us now be more specific for the computations of  ${\bf S}  ^{[\lambda , j ]} (t, X)$, ${\bf B}  ^{[ j ]} (x, t, X)$
 and  ${\bf E}  ^{[ j ]} (x, t, X)$ with $j\leq 1$, in particular for $X=0$.

 {\it First term ($j=0$) for the electromagnetic field.}
 
We have  ${\bf B}  ^{[ 0 ]} (x, t, X) = {\bf B}  ^{free} (x, t, X)$ and ${\bf E}  ^{[ 0 ]} (x, t, X) = {\bf E}  ^{free} (x, t, X)$. In particular,
if $X=0$ then ${\bf B}  ^{[ 0 ]} (x, t, 0) = {\bf E}  ^{[ 0 ]} (x, t, 0)= 0$.

{\it First term ($j=0$) for the spin.} 

 The first term for the spin satisfies an equation similar to the Bloch equation, namely,
\be\label{MaxSym-5} {d \over dt} {\bf S}  ^{[\lambda, 0 ]} (t, X) = 2 ( {\bf B}^{ext} + {\bf B}^{[0]}  (x_{\lambda} , t, X) )
\times  {\bf S}  ^{[\lambda , 0]} (t, X)\ee
with $  {\bf S}  ^{[\lambda, 0 ]} (0, X) = ( \sigma_1^{[\lambda]} , \sigma_2^{[\lambda]} , \sigma_3^{[\lambda]} ) $.
Thus, if $X=0$ and when ${\bf B}^{ext} = (0, 0, |B|)$,  in the case of a unique particle fixed at the origin (we then omit the index $\lambda$ in that case), we have,
\be\label{S-0}{\bf S}^{[0]}  (t, 0) = \Big ( \cos ( 2 |B|t) \sigma _1 -  \sin ( 2 |B|t) \sigma _2 \ , \
\sin ( 2 |B|t) \sigma _1 +  \cos ( 2 |B|t) \sigma _2 \ , \ \sigma_3 \Big ).  \ee

{\it Second term ($j=1$) for the fields.} 

 The functions $(x , t) \rightarrow {\bf B} ^{[1]}  (x, t, X)$ and $(x , t) \rightarrow {\bf E} ^{[1]}  (x, t, X)$
are solutions to Maxwell equations with a entirely vanishing initial data, together with a vanishing charge density and a current density given in formula   (40) of  \cite{A-J-N} by,
 $$ {\bf J}^{[1]} (x , t, X) =  \sum _{\lambda = 1} ^N {\bf S^{[\lambda , 0]}}
(t, X) \times  {\rm grad }
\rho (x-x_{\lambda}),$$
$$
\rho (x) = (2 \pi )^{-3}  \int _{\R^3} |\chi (|k|)|^2  \ \cos (k \cdot  x) dk,$$
where $\chi$  is the function appearing in  (\ref{7.3}). There exists in particular a function 
${\bf A}^{[1]} (x , t, X)$ taking values in  $( {\cal L} ({H}_{sp}))^3$ and satisfying,
\be\label{champ}{\bf B} ^{[1]}  (x, t, X) = {\rm curl}\ {\bf A}^{[1]} (x , t, X),
\qquad {\bf E} ^{[1]}  (x, t, X) = - \frac {\partial } {\partial t}
{\bf A}^{[1]} (x , t, X) \ee
$$ \left ( \frac {\partial^2 } {\partial t^2 } - \Delta _x \right )
{\bf A}^{[1]} (x , t, X) =  {\bf J}^{[1]} (x , t, X)$$
and ${\bf A}^{[1]} (x , 0, X) = \partial _t {\bf A}^{[1]} (x , 0, X) = 0$.
Still in the case of a unique  particle fixed at the origin (omitting $\lambda$ from the notation and setting $x_{\lambda}= 0$), assuming
${\bf B}^{ext} = (0, 0, |B|)$ and in the case
 $X = 0$, we can write   ${\bf J}^{[1]} (x , t, 0) = \sum _{j= 1}^3
{\bf G}^{[j]} (x , t) \sigma _j$, where the  ${\bf G}^{[j]} $ are all real valued and satisfy according to
 (\ref{S-0}),
$$ ( {\bf G}^{[1]} + i {\bf G}^{[2]}  )(x , t) =
\begin{pmatrix}  i \frac {\partial \rho} {\partial x_3}  \\   - \frac {\partial \rho} {\partial x_3}  \\
\frac {\partial \rho} {\partial x_2}  - i\frac {\partial \rho} {\partial x_1}   \end{pmatrix} e^{-2i|B|t},\qquad
{\bf G}^{[3]} (x , t) = \begin{pmatrix} -  \frac {\partial \rho} {\partial x_2} \\
 \frac {\partial \rho} {\partial x_1} \\ 0 \end{pmatrix}. $$
For every real number $\omega$, let 
 \be\label{def-u} u(x , t, \omega ) = (2 \pi )^{-3}  \int _{\R^3 \times [0, t] } |\chi (|k|)|^2  \ \cos (k \cdot  x)
 \ \frac { \sin (|k|(t-s)) } {|k|}  e^{-i\omega s} dk\ ds. \ee
One has,
$$ \left ( \frac {\partial ^2 } {\partial t^2} - \Delta \right )  u(x, t, \omega) =
e^{-i\omega t} \rho (x) $$
and  $u(x, 0, \omega) = \partial _t u(x , 0, \omega ) = 0$.
We then can write ${\bf A}^{[1]} (x , t, 0) = \sum _{j= 1}^3
{\bf F}^{[j]} (x , t) \sigma _j$ where all the  ${\bf F}^{[j]} $are real-valued and
$$  ( {\bf F}^{[1]} + i {\bf F}^{[2]}  ) (x, t) =
\begin{pmatrix} i \frac {\partial u} {\partial x_3}  \\   - \frac {\partial u} {\partial x_3}  \\
\frac {\partial u} {\partial x_2}  - i\frac {\partial u} {\partial x_1} \end{pmatrix}(x, t, 2 |B|),\qquad
{\bf F}^{[3]} (x , t) = \begin{pmatrix} -  \frac {\partial u} {\partial x_2} \\
 \frac {\partial u} {\partial x_1} \\ 0 \end{pmatrix} (x, t, 0).$$
The fields  ${\bf B} ^{[1]}  (x, t, 0)$ and ${\bf E} ^{[1]}  (x, t, 0)$ can then be deduced using (\ref{champ}) 
when there is a unique particle fixed at the origin and when $X=0$. Since the function $ u(x, t, \omega) $ defined in (\ref{def-u}) is radial,
one has  $\partial _j \partial _k u(0, t, \omega) =
(1/3) \delta _{jk} \Delta u(0, t, \omega)  $. Consequently, for $x=0$,
\be\label{B-1} {\bf B} ^{[1]}  (0, t, 0)= \frac {2} {3}  \begin{pmatrix}
\Delta \ {\rm Re}\ u \\
- \Delta \ {\rm Im}\ u \\
0  \end{pmatrix} (0, t, 2|B|) \ \sigma _1 +
 \frac {2} {3}  \begin{pmatrix}
\Delta \ {\rm Im}\ u \\
\Delta \ {\rm Re}\ u \\
0  \end{pmatrix} (0, t, 2|B|) \ \sigma _2 +
\frac {2} {3}  \begin{pmatrix}
0 \\ 0 \\
\Delta u \end{pmatrix} (0, t, 0) \ \sigma _3.\ee

{\it Second term ($j=1$) for the spin symbol.}  

In the aim to write the differential system satisfied by  ${\bf S}  ^{[\lambda , 1 ]} (t, X)$ we use the first two terms of the Mizrahi series. For any suitable $F$ and $G$  defined on $H^2$ taking possibly values  in ${\cal L}({\cal H}_{sp})$, we set,
$$ C_0 (F , G) (X) = F(X) G(X),\qquad
C_1 (F , G) (X) = (1/2)  \sum _{j }
\left ( \frac { \partial F} {\partial q_j} - i \frac { \partial F} {\partial p_j}\right ) \
\left ( \frac { \partial G} {\partial q_j} + i \frac { \partial G} {\partial p_j}\right ). $$
If ${\bf F}$  and ${\bf G}$ take values in  $({\cal L}({\cal H}_{sp}))^3$, we define $ C_0 ^{\times , sym} ({\bf F} , {\bf G})$ and $ C_1 ^{\times , sym} ({\bf F} , {\bf G})$
as functions on $H^2$, taking values in $({\cal L}({\cal H}_{sp}))^3$ such that, for example,
$$  C_j ^{\times , sym} ({\bf F} , {\bf G})_3 = \frac {1} {2} \Big [C_j ( F_1 , G_2) + C_j (G_2, F_1)
- C_j (F_2, G_1) - C_j ( G_1, F_2) \Big ].$$
With these notations, one shows in 
   \cite{A-J-N}, (formulas (95) and (100)) that, for every $X$ in $H^2$,
\be\label{9-4} \frac  {d } { dt} {\bf S}  ^{[\lambda , 1 ]} (t, X ) =
2 ( {\bf B} ^{ext}  + {\bf B} ^{[0]}  (x_{\lambda} , t, X) )
\times  {\bf S}  ^{[\lambda , 1 ]} (t, X) +
2 C_0^{\times , sym} ({\bf B} ^{[1]}( x_{\lambda}, t, \cdot)  , {\bf S} ^{[\lambda , 0]} (t, \cdot ))(X) + ... \ee
$$ ... + 2 C_1^{\times , sym} ({\bf B} ^{[0]} ( x_{\lambda}, t, \cdot) , {\bf S} ^{[\lambda , 0]} (t, \cdot ))(X) .$$
We shall make explicit this  computation in the case of a unique particle fixed at the origin when ${\bf B} ^{ext} = (0, 0, |B|)$ and if $X=0$. Recall that in that case  we omit the index  $\lambda$. 
The function ${\bf S}  ^{[ 1 ]} (t, X)$ taking values in  $({\cal  L} ({\cal H}_{sp}))^3 = ({\cal  L} ({\bf C}^2))^3$, selfadjoint,
can be expressed as,
\be\label{def-F} {\bf S}  ^{[ 1 ]} (t, X) = {\bf F}^{[0]} (t, X) I + \sum _{j=1}^3 {\bf F}^{[j]} (t, X)
\sigma _j .\ee
Since our issues is to especially study $< {\bf S}  ^{[ 1 ]} (t, 0) a, a>$ with $a =(1, 0)$ then
we only make explicit ${\bf F}^{(0)} (t, 0)$ and ${\bf F}^{(3)} (t, 0)$.
For that purpose, we shall make explicit the first two terms of (\ref{9-4}).

Since $\sigma _j \sigma _k  + \sigma _k \sigma _j = 2 \delta _{jk} I $ we deduce from (\ref{S-0}) and (\ref{B-1}) that,
$$ C_0^{\times , sym} (  {\bf B} ^{[1]}  (0 , t, \cdot ) \ , \   {\bf S}  ^{[ 0 ]} (t, \cdot) ) (0) =
{\bf G} ^{[0]}(t) I ,\qquad {\bf G} ^{[0]}(t) = \Big ( 0, 0, \Phi_0 (t) \Big ) $$
with
$$ \Phi_0 (t) =  \frac {4} {3}  \cos (2|B|t)\  \Delta ( {\rm Im}\  u) (0, t, 2|B|) + \frac {4} {3}
 \sin (2|B|t)\  \Delta ({\rm Re} \ u) (0, t, 2|B|), $$
where $u(x, t, \omega)$ is the function defined in (\ref{def-u}). As a consequence,
\be\label{Phi-0} \Phi_0 (t) = -  \frac {4} {3} (2 \pi )^{-3}  \int _{\R^3 \times [0, t]}
|\chi (|k|)|^2 \ |k| \  \sin (|k|(t-s))   \sin (2|B|(t-s))  dk\ ds.  \ee
Next, let us make explicit, still omitting the index $\lambda$ from the notation, the function
$C_1^{\times , sym} ({\bf B} ^{[0]} ( 0, t, \cdot) , {\bf S} ^{[ 0]} (t, \cdot )(0) $.
For every suitable functions $F$ and  $G$  defined on $H^2$ and taking values in ${\cal L}({\cal H}_{sp})$ while assuming that one the two functions is proportional to identity, we note that, 
 $$  C_1 (F , G) + C_1 (G , F) =
  \sum _{j } \frac { \partial F} {\partial q_j}\
  \frac { \partial G} {\partial q_j} \ + \
 \frac { \partial F} {\partial p_j}\
  \frac { \partial G} {\partial p_j}.$$
Since the  $ B_j ^{[0]} ( 0, t, X)$ are scalar elements (that is to say, elements of  ${\cal L} ({\cal H}_{sp})$ proportional to identity) and in view of 
$ B_j ^{[0]} ( 0, t, X) = B_{j, 0, t} \cdot X$, we see for suitable functions
$F$ that,
$$   \Big ( C_1 (F ,  B_j ^{[0]} ( 0, t, \cdot)) + C_1 ( B_j ^{[0]} ( 0, t, \cdot)  , F) \Big )
 (X) =  dF(X) ( B_{j, 0, t} ). $$
One can write,
 $$ C_1^{\times , sym} ({\bf B} ^{[0]} ( 0, t, \cdot) , {\bf S} ^{[ 0]} (t, \cdot ))(0) =
 \sum _{j=1}^3 {\bf G} ^{[j]}(t) \sigma _j $$
 and for any $V$ in $H^2$,
 $$ d {\bf S} ^{[ 0]} ( t, 0 ) (V) =\sum _{j=1}^3 {\bf H} ^{[ j]} (t, V) \sigma _j$$
 $$ 2 {\bf B^{[0]}} (0, t, V) \times  {\bf S} ^{[ 0]} ( t, 0 ) = \sum _{j=1}^3
 {\bf K}  ^{[ j]} (t, V) \sigma _j.$$
Let us write the existing relations between these functions. From the foregoing, it is clear from that, 
 $$  {\bf G} ^{[j]}(t)  =  \frac {1}  {2}  \begin{pmatrix}
 H_3  ^{[j]} (t, B_{2, 0, t}) - H_2  ^{[j]} (t, B_{3, 0, t})\\
  H_1  ^{[j]} (t, B_{3, 0, t}) - H_3  ^{[j]} (t, B_{1, 0, t})\\
  H_2  ^{[j]} (t, B_{1, 0, t}) - H_1  ^{[j]} (t, B_{2, 0, t})\end{pmatrix}.$$
Differentiating the system (\ref{MaxSym-5}) at point $X= 0$ with $x_{\lambda} =0$ and observing that 
${\bf B} ^{[0]} ( 0, t, 0) = 0$ we have for every $V$ in $H^2$,
$$  \frac  {d } { dt} d {\bf S} ^{[ 0]} ( t, 0 ) (V) =
2 {\bf B^{ext}} \times    d {\bf S} ^{[ 0]} ( t, 0 ) (V)
+ 2 {\bf B^{[0]}} (0, t, V) \times  {\bf S} ^{[ 0]} ( t, 0 ). $$
We have above used the fact that the symbol $B_j ^{[ 0]} ( 0, s, V )$  depends linearly on  $V$. As $S_j (0, h) = I \otimes \sigma_j$, the Wick symbol  $S_j ^{[ 0]} ( 0, X )$
is independent on $X$ and consequently $d {\bf S} ^{[ 0]} ( 0, 0 ) (V) = 0$.
Therefore,
$$  \frac  {d } { dt} {\bf H} ^{[ j]} (t, V) =
2 {\bf B^{ext}} \times  {\bf H} ^{[ j]} (t, V) +  {\bf K}  ^{[ j]} (t, V)$$
and ${\bf H} ^{[ j]} (0, V) = 0$. With $ {\bf B^{ext}} = (0, 0, |B|)$ one gets,
$$ {\bf H} ^{[ j]} (t, V)  = \int _0^t   \begin{pmatrix}
 \cos (2|B| (t-s))K_1^{[ j]}  (s, V)  -  \sin  (2|B| (t-s)) K_2^{[ j]}  (s, V)
  \\
\sin (2|B| (t-s))K_1^{[ j]}  (s, V) +  \cos  (2|B| (t-s))K_2^{[ j]}  (s, V) \\
K_3^{[ j]}  (s, V)  \end{pmatrix} ds.  $$
According to (\ref{S-0}) and using $B_1 ^{[ 0]} ( 0, s, V ) = B_{1, 0, s} \cdot  V$ one obtains,
$${\bf K} ^{[3]}(t, V) =  2 \begin{pmatrix} B_{2, 0, t} \cdot V \\
- B_{1, 0, t} \cdot V  \\ 0 \end{pmatrix}.$$
In view of (\ref{7.3}) and  (\ref{Bjxt}),
$$ B_{mxt} \cdot  B_{nxs} = 0 \quad {\rm if} \quad m\not = n,
\qquad B_{mxt} \cdot  B_{mxs} = \frac {2} {3} (2 \pi )^{-3}
\int_{\R^3} |\chi(|k|)|^2  |k| \cos ( |k| (t-s)) dk. $$
Consequently ${\bf G} ^{[3]}(t) = (0, 0, \Phi_3 (t))$ with,
$$ \Phi_3 (t) =  - 2 \int _0^t \cos ( 2|B| (t-s)) B_{1,0,t} \cdot  B_{1,0,s}
ds$$
and then
\be\label{Phi-3} \Phi_3 (t) = - \frac {4} {3} (2 \pi )^{-3}
\int_{\R^3 \times [0, t]} |\chi(|k|)|^2  |k|
\cos ( |k| (t-s))  \cos (2|B| (t-s)) dk ds.\ee
As ${\bf B} ^{[0]}  (x , t, 0)= 0$ the system (\ref{9-4}) gives for the functions  ${\bf F} ^{[j]}(t)$ defined in (\ref{def-F}) when there is only one particle fixed at the origin and assuming ${\bf B} ^{ext} = (0, 0, |B|)$, in the case  $X=0$,
$$ \frac {d} {dt}  {\bf F} ^{[j]}(t) = 2 {\bf B} ^{ext} \times   {\bf F} ^{[j]}(t) +
2 {\bf G} ^{[j]}(t),\qquad 0 \leq j \leq 3$$
and $ {\bf F} ^{[j]}(0)= 0$. As ${\bf G} ^{[j]}(t)= (0, 0, \Phi_j (t))$ ($j=0$
and $j=3$), one then deduces that,
$$  {\bf F} ^{[j]}(t) = \left ( 0, 0, \ 2\int _0^t \Phi_j (s) ds \right ),\qquad j=0 \ \ {\rm  and} \ \ j=3.$$

{\it Application to the average values.} 

Consider a unique particle fixed at the origin in a constant magnetic field ${\bf B} ^{ext} = (0, 0, |B|)$ and assume that $X=0$. We shall now make explicit 
$<{\bf S}  ^{[ 1 ]} (t, 0)a , a>$ for $a = (1, 0)$. According to the foregoing facts,
$$<{\bf S}  ^{[ 1 ]} (t, 0)a , a> = \left ( 0, 0, \ 2\int _0^t ( \Phi_0 (s)+ \Phi_3 (s)) ds \right ).  $$
Then taking  (\ref{Phi-0}) and (\ref{Phi-3}) into account, 
$$ \Phi_0 (t)+ \Phi_3 (t) = - \frac {4} {3} (2 \pi )^{-3}
\int_{\R^3 \times [0, t]} |\chi(|k|)|^2  |k| \cos ( (|k| - 2 |B|) s) ds dk$$
$$ = - \frac {4} {3} (2 \pi )^{-3}
\int_{\R^3 } |\chi(|k|)|^2  |k| \frac { \sin ( (|k| - 2 |B|) t)} {|k| - 2 |B|} dk. $$
We then have proved the following expression,
$$< S_3  ^{[ 1 ]} (t, 0)a , a> =  \frac {8} {3} (2 \pi )^{-3}
\int_{\R^3 } |\chi(|k|)|^2  |k| \frac {   \cos ( (|k| - 2 |B|) t) - 1}
{(|k| - 2 |B|)^2} dk. $$
The above formula amounts for the second term
 (coefficient of $\hslash$)
of the asymptotic expansion of the spin third component average when the initial data is taken as $\Psi_{0, h} \otimes a$ with  $a= (0, 1)$. As a consequence, according to the model coming from QED and contrarily to the Bloch model, the spin cannot stay eternally fixed at the position $(0, 0, 1)$. Let us mention that 
 H\"ubner and H. Spohn in \cite{H-S} prove for a related model that, the spin ${\bf S}  ( t, X, \hslash)$ tends, as  $t$ goes to infinity, to a limit being at a distance ${\cal O} (h)$ of the point   $(0, 0, -1)$.

\section{Approximate evolution of  photon number.}\label{s3}

The average photon number taken in a coherent state $\Psi_{X  \hslash}$ is $|X|^2/2 \hslash$. It consequently goes to infinity as $ \hslash$ tends toward zero. Nevertheless, if the initial state is taken as $\Psi_{X  \hslash} \otimes a$ where  $a$ in ${\cal H}_{sp}$ has unit norm then we shall prove that the photon number variation between times $0$ and $t$ has limit when  $ \hslash$ tends to zero.
As a matter of fact, a radiated or absorbed photon semiclassical analysis can be carried out.

The photon average number at time $t$ assuming that the initial state is 
$\Psi_{X  \hslash} \otimes a$ has the value  $< N(t, X  \hslash) a , a>$ where $N(t, X ,  \hslash)$ 
is an operator acting in ${\cal H}_{sp}$ being the Wick symbol of the observable  $N(t, h)$. It is derived in \cite{A-J-N} that we can write,
$$ {d \over dt } N( t, X,  \hslash) = \sum _{r=0}^M \hslash^r
 N^{[r]}  (t, X)+ \hslash^{M+1} R_M (t, X, \hslash)$$
where the $ N^{[r]} (t, X)$ are explicitly computable functions 
 and  where $R_M (t, \cdot , h)$ belongs to a class $S(H^2, K Q_t)$, with
 $K$ being some constant and where the norm of $R_M (t, \cdot , \hslash)$
 in this class is bounded independently of $t$ belonging to a compact set of $\R$
 and of $\hslash$ lying in $(0, 1]$.

Let us now recall the first term  $ N^{[0]} (t, X)$. In the general case where $X$ is an arbitrary element of  $H^2$, the coherent state  $\Psi_{X h}$ has neither definite frequency nor definite polarization. 
We shall then associate with it two specific fields ${\bf B}^{pol} (x , X)$ and ${\bf E}^{pol} (x , X)$ having no counterpart in classical physics. If the photon had a definite circular polarization then these two fields would simply be the magnetic and electric fields with a possibly further change of sign according to the polarization. 

{\it Polarized fields.} 

For every $k$ in $\R^3 \setminus \{ 0 \}$ the space
$E_+(k) $ stands for the set of all $(q, p) \in \R^6$ satisfying
$q\cdot k = p \cdot k = 0$ and
$ k \times  q = -p |k|$.  Likewise, we denote by   $E_-(k)$ the set of all $(q, p) \in \R^6$
verifying $q\cdot k = p \cdot k = 0$ and
$ k \times  q = p |k| $. Accordingly, any point  $(q, p) \in \R^6$ being such that 
$q\cdot k = p \cdot k = 0$ has now a unique decomposition written as $(q , p) = \Pi_+ (k) (q , p) +  \Pi_- (k) (q , p)$ where $\Pi_{\pm} (k) (q , p)$ belongs to $E_{\pm}(k) $. For any  $X$ in
$H^2$ we denote by $\Pi_{\pm}X$  the element in $H^2$ verifying
$( \Pi_{\pm}X) (k) = \Pi_{\pm} ( X(k))$ for a.e. $k\in \R^3 \setminus \{ 0 \}$.
For all $x$ in $\R^3$, denoting by $B_{jx}$ the element in  $H^2$ defined in (\ref{7.3}), we define a function  $ B_{j}^{pol} (x , \cdot)$ on  $H^2$ and an unbounded operator $  B_{j}^{pol}(x)$ in ${\cal H}_{ph}$ by,
$$B_{j}^{pol} (x , X)  = \Big ( (\Pi_+ - \Pi_-) B_{jx}\Big ) \cdot X,\qquad
  B_{j}^{pol}(x) = \Phi_S \Big ( (\Pi_+ - \Pi_-) B_{jx}\Big ).$$
We define likewise  $ E_{j}^{pol} (x , \cdot)$  and $  E_{j}^{pol}(x)$.
For all $t$ in $\R$, we set,
$$B_{j}^{pol, free} (x , t,  X)  = \Big ( (\Pi_+ - \Pi_-) B_{jx}\Big ) \cdot
\chi_t (X),\qquad E_{j}^{pol, free} (x , t,  X)  = \Big ( (\Pi_+ - \Pi_-) E_{jx}\Big ) \cdot
\chi_t (X). $$
It is shown in \cite{A-J-N} that,
\be\label{evol-phot}  N^{[ 0 ]} (t, X) = -\sum _{\lambda = 1}^N \sum _{j=1}^3
  E_j^{pol, free} (x_{\lambda}, t, X ) \
S_j^{[\lambda , 0 ]}  ( t,X ). \ee
Accordingly, if the initial coherent state $\Psi_{X ,\hslash}$ has a well definite circular polarization, that is to say, if   $X(k)$ belongs to $E_{\pm }(k) $ for a.e. $k$
in $\R^3 \setminus \{ 0 \}$, then we obtain a semiclassical estimation of the emitted or absorbed photon number. Namely,
\be\label{evol-phot-pol} \frac {d} {dt}  N( t, X,  \hslash) =
\mp  \sum _{\lambda = 1}^N  {\bf E}^{free} (x_{\lambda}, t, X ) \cdot
{\bf S} ^{[\lambda , 0 ]}  ( t,X ) + {\cal O} (\hslash).  \ee
Let us now emphasize that this equality is consistent with the energy conservation.

From  (\ref{7.3}) - (\ref{def-elec}) and (\ref{chi-t}), we have, when  $X= (q , p)$
is identified with $q+ip$,
$$  B_j^{free} (x, t, X ) = \int _{\R^3} \frac { \chi (|k|) |k|^{1/2}}
{ (2\pi) ^{3/2} } \Big [ \cos ( k\cdot x - t |k|)
 \frac { (p(k) \times  k) \cdot e_j } {|k|} +
\sin  ( k\cdot x - t |k|)  \frac { (q(k) \times  k) \cdot e_j } {|k|} \Big ] dk. $$
If $X(k)$ belongs to $E_{\pm }(k) $ for a.e. $k$
in $\R^3 \setminus \{ 0 \}$, we see that  $ k \times  q(k)= \mp |k| p(k)$ and
$ k \times  p(k)= \pm  |k| q(k)$. Therefore,
$$  B_j^{free} (x, t, X ) = \pm  \int _{\R^3} \frac { \chi (|k|) |k|^{1/2}}
{ (2\pi) ^{3/2} } \Big [- \cos ( k\cdot x - t |k|)
 q_j(k) +
\sin  ( k\cdot x - t |k|) p_j(k) \Big ] dk $$
and
$$ \frac {d} {dt}  B_j^{free} (x, t, X ) = \mp  \int _{\R^3} \frac { \chi (|k|) |k|^{1/2}}
{ (2\pi) ^{3/2} }\ |k| \  \Big [  \sin ( k\cdot x - t |k|)
 q_j(k) + \cos ( k\cdot x - t |k|)   p_j(k) \Big ] dk. $$
Besides, in view of  (\ref{7.3}) - (\ref{def-elec}) together with (\ref{chi-t}), we have
$$  E_j^{free} (x, t, X ) = -\int _{\R^3} \frac { \chi (|k|) |k|^{1/2}}
{ (2\pi) ^{3/2} } \Big [ \sin ( k\cdot x - t |k|) q_j (k) + \cos ( k\cdot x - t |k|)
p_j (k) \Big ] dk.$$
A coherent state $\Psi _{X , h}$ never has an exactly definite frequency. In order to formulate the idea that it is approximatively true for a frequency  $\nu$ we can consider an element 
 $X = (q , p)$ of $H^2$ where $X(k) = 0$ excepted if $||k| - \nu | < \varepsilon$ with $\varepsilon >0$ being a very small parameter. In that condition, if in addition 
$X(k)$ belongs to $E_{\pm }(k) $ for a.e. $k$, then
$$ \frac {d} {dt}  B_j^{free} (x, t, X ) = \pm \nu  E_j^{free} (x, t, X ) + {\cal O}(\varepsilon) .$$
For that reason, under the same assumptions,
$$  \frac {d} {dt}  N( t, X,  \hslash) = - \frac {1} {\nu }
\sum _{\lambda = 1}^N
{\bf S} ^{[\lambda , 0 ]}  ( t,X ) \cdot  \frac {d} {dt}
{\bf B} ^{free} (x_{\lambda}, t, X ) + {\cal O} (\varepsilon )
+ {\cal O}(\hslash).$$
Then, taking into account (\ref{MaxSym-5}), we deduce,
$$  \frac {d} {dt} \left [  N( t, X,  \hslash)  +
\frac {1} {\nu } \sum _{\lambda = 1}^N
{\bf S} ^{[\lambda , 0 ]}  ( t,X ) \cdot
({\bf B} ^{ext} +  {\bf B} ^{free} (x_{\lambda}, t, X ))  \right ] = {\cal O}  (\varepsilon )
+ {\cal O}(\hslash),$$
which seems physically realistic.

\section{Transition probabilities.}\label{s4}

When $X$ and  $Z$ are elements of $H^2$,  $a$ and $b$
two elements of   ${\cal H}_{sp}$ both with unit norm, we shall show that the transition probabilities $ | < e^{i{t \over \hslash}  H(\hslash)} ( \Psi _{X \hslash}\otimes a )  , \Psi _{Z \hslash}\otimes b > | $ is very low excepted if   $Z$ is closed to $\chi_t(X)$,
where $\chi_t$ is defined in (\ref{chi-t}). More specifically, we shall prove that,
\be\label{1.3}
 | < e^{i{t \over \hslash}  H(\hslash)} ( \Psi _{X h}\otimes a )  , \Psi _{Z \hslash}\otimes b > | \leq
  e^{ \frac {1} {2} Q_t( {\cal F} (X-\chi_t^{-1} (Z)  )) ^{1/2}}   e^{-{1\over 4\hslash}|X-\chi_t^{-1} (Z) |^2 }
  \ee
where $\chi_t$ is the symplectic map in $H^2$ defined in (\ref{chi-t}) with $Q_t$ defined in  (\ref{def-2-Qt} ) and  ${\cal F}(q , p) = (-p, q)$.

We first define  $U_h^{red} (t)$ the operator,
$$  e^{-i{t \over\hslash}  H(\hslash)} = e^{-i{t \over\hslash} (H_{ph} \otimes I) } U_h^{red} (t). $$
According to Theorem 7.1  of \cite{A-L-N-1} and Theorem 1.4 (point iv))  in \cite{A-J-N},
the Wick symbol of the operator $U_h^{red} (t)$ belongs to the class of symbols  $S(H^2 ,  Q_t)$ with norm one and taking values in  ${\cal L} ( {\cal H}_{sp})$. Taylor formula and estimates (\ref{1.bis}) satisfied by this Wick symbol show that it has an holomorphic extension  $\Phi_t$ in $H_{\bf C}^2$ taking values in 
${\cal L} ( {\cal H}_{sp})$ and verifying,
$$ |\Phi_t (Z)|\leq \ e^{ Q_t( {\rm Im} Z) ^{1/2}},\qquad
Z\in (H_{\bf C})^2.$$
We then know that, for any $X = (q , p)$ and  $Y = (q' , p')$
in $H^2$, for every $a$ and  $b$ in ${\cal H}_{sp}$,
\be\label{7.5-1} \frac { <U_h^{red} (t) \Psi_{q,p, h}\otimes a , \Psi_{q',p', h}\otimes b  >  }
{ < \Psi_{q,p, h} , \Psi_{q',p', h}> }
=  \Bigg < \Phi_t
 \left ( { q+ip \over 2} +  { q' -ip' \over 2} ,  { p-iq \over 2} +  { p'+iq' \over 2}
 \right )a , b \Bigg > .\ee
As a matter of fact,  the left hand side is an holomorphic function in $q+ip$ and antiholomorphic
in $q' + i p'$ with a restriction to the diagonal equaling  to $<U_h^{red} (t) \Psi_{q,p, \hslash}\otimes a , \Psi_{q,p,\hslash}\otimes b  > $.  We also know that,
$$ | < \Psi_{X, h} , \Psi_{Y, h}>| \leq e^{-{1\over 4\hslash}|X-Y|^2 }. $$
As a consequence, for $a$ and $b$ with unit norm,
$$ \Big  |  <U_h^{red} (t) \Psi_{X, \hslash}\otimes a , \Psi_{Y, \hslash}\otimes b  > \Big |
\leq   e^{ \frac {1} {2} Q_t( {\cal F} (X-Y)) ^{1/2}}   e^{-{1\over 4\hslash}|X-Y|^2 }. $$
 One then gets,
$$  < e^{-i{t \over h}  H(h)} (\Psi_{X , \hslash} \otimes a ), (\Psi_{Z  , \hslash} \otimes b )>  =
< e^{-i{t \over \hslash} (H_{ph} \otimes I) } U_h^{red} (t) (\Psi_{X , \hslash} \otimes a ), (\Psi_{Z  , \hslash} \otimes b )> $$
$$ = <  U_h^{red} (t) (\Psi_{X , \hslash} \otimes a ), (\Psi_{\chi_{-t} (Z)  ,\hslash} \otimes b )>. $$
We thus deduce (\ref{1.3}).

\medskip

laurent.amour@univ-reims.fr\newline
LMR EA 4535 and FR CNRS 3399, Universit\'e de Reims Champagne-Ardenne,
 Moulin de la Housse, BP 1039,
 51687 REIMS Cedex 2, France.

lisette.jager@univ-reims.fr\newline
LMR EA 4535 and FR CNRS 3399, Universit\'e de Reims Champagne-Ardenne,
 Moulin de la Housse, BP 1039,
 51687 REIMS Cedex 2, France.

jean.nourrigat@univ-reims.fr\newline
LMR EA 4535 and FR CNRS 3399, Universit\'e de Reims Champagne-Ardenne,
 Moulin de la Housse, BP 1039,
 51687 REIMS Cedex 2, France.

\end{document}